\documentstyle[11pt,epic,eepic]{article}

\def\beq{\begin{equation}}
\def\eeq{\end{equation}}
\def\bea{\begin{eqnarray}}
\def\eea{\end{eqnarray}}

\def\nn{\nonumber}
\def\ox{\otimes}

\title
{\LARGE\bf The Effects of a Vector-like Doublet \\
as The Fourth Generation to $R_b$}
\bigskip
\bigskip
\vspace{1cm}
\author{ Tadashi YOSHIKAWA\thanks{e-mail:
yosikawa@theo.phys.sci.hiroshima-u.ac.jp}  \\
                             \\
\normalsize \em Department of Physics, Hiroshima University\\
\normalsize \em 1-3-1 Kagamiyama, Higashi-Hiroshima, 739\\
\normalsize \em Japan}

\date{}
\begin{document}
\setlength{\baselineskip}{24pt}
\maketitle
\begin{picture}(0,0)
\put(325,240){HUPD-9528}
\put(325,225){November, 1995}
\end{picture}
\vspace{-24pt}

\vspace{3cm}
\begin{abstract} \large
We study the effects of a vector-like SU(2) doublet quarks
  as the fourth
generation
on  $R_b$ and $R_c$.
 By considering  the constraint from the experimental data
of forward-backward asymmetry, $A_{FB}^b$ and $A_{FB}^c$,as
well as
 FCNC among  light quarks,  we show that there is an allowed
region of parameters for  $R_b$ but not for $R_c$.
\end{abstract}
\newpage
\section{Introduction}
The standard model(SM) is consistent with almost all of the
experiments.  Most of the experiments are consistent with
the predictions of
 SM with sufficient accuracy.
However, the recent experimental data at LEP shows that there are the
deviations from the standard model prediction
in the Z boson partial width ratios
$R_b =\Gamma[Z \to b {\bar b}] / \Gamma[Z
\to hadrons]$
and $R_c$. The experimental values $R_b = 0.2219 \pm 0.0017$ and
$R_c = 0.1540 \pm 0.0074$ have already been different from the values
$R_b^{SM} = 0.2157$ and $R_c^{SM} = 0.1722$ which are predicted by SM with top
quark mass $m_t = 175GeV$\cite{hagi}. If these
deviations show the existence of new
physics, we must examine all of the possibility
of beyond SM.

At present, the predictions for $R_b$ and $R_c$ in many
beyond the standard models are compared with the
experiments, any of them can not explain the deviations
\cite{civpro}.
In minimal supersymmetric model, in order to explain the deviation of
the $R_b$ there must be light
chargino and top squark\cite{WKK}.
In the
extended technicolor model(ETC),
in the works \cite{wu} it is shown that the diagonal
extended technicolor(ETC) interaction may solve the
problem.
However, the effect contribute to $T$ parameter, too\cite{yosi3}.
If the contribution of the
diagonal interaction to $Zbb$ vertex
is large enough to cancel the other corrections
for the $Zbb$ vertex, $T$ is unacceptably large.

In those
 works mentioned above, the corrections to  $Zbb$ vertex from loop contribution
of new particles
 are  mainly studied.
 In this paper,
 we consider the shift of Z coupling at tree
level  in the context of the non-trivial extension
of quark sector into the forth generation.
The shift comes from the flavor mixing
like CKM\cite{CKM} matrix.
In this work, we discuss the effect of
a vector-like SU(2)doublet quarks as the fourth generation to the $Zbb$
vertex.
In this model
the right-handed Z coupling differs from that of the SM.
Hence, the Z
decay widths are  different from  those of  the SM at tree level. The effects
of the vector-like doublet quarks as the third generation,
 were already studied in
Ref.\cite{ple}.  However, because  they assume that the  vector
like  quarks are the third
generation,
their model can not satisfy the constraint from the
forward-backward asymmetry, $A_{FB}^b$.(See section 2.)
Therefore, we study a vector-like doublet quarks as the forth generation.

This paper is developed as following. In section 2, we
discuss  the model including a vector-like doublets as the
forth generation. In section 3, the effects on
$R_b$ and $R_c$ is shown in  the simple case that the  flavor mixing occurs
only between the third and the forth generation.
In section 4, we show the relation between the mixing angle
and mass of new quarks.  Section 5 is devoted to the
conclusion.

\section{The Model}
First, we study the effects of new fermions which are
 transformed in a vector-like way
under the electro-weak symmetry $SU(2)_L \ox U(1)_Y$.
 The SU(2) and hypercharge of ordinary
fermions $Q$
and new vector-like fermions $Q^\prime$ are assigned in  the following
way.
\bea
  Q_L^i =  \pmatrix{
             u^i  \cr
             d^i
                    }_L \sim ( 2, \frac{1}{3} ),
  u_R^i \sim ( 1, \frac{4}{3}),
  d_R^i \sim ( 1, - \frac{2}{3}), \nn
\eea
\bea
  Q_L^\prime = \pmatrix{
                 u^\prime \cr
                 d^\prime
                        }_L \sim ( 2, \frac{1}{3}),\nn \\
  Q_R^\prime = \pmatrix{
                 u^\prime \cr
                 d^\prime
                        }_R \sim ( 2, \frac{1}{3}). \nn
\eea
Because the vector-like fermions   are SU(2)
doublet,  the left-handed and right-handed
chiralities have  the same charge under
$SU(2)_L$ transformation.

The neutral current interactions of fermions can be written as
\bea
{\cal L} = - \frac{g}{2 cos\theta_w}
                   \sum_i [ L_i \bar{Q}_L^i
                                    \gamma^\mu
                                       Q_L^i
                           + R_i \bar{Q}_R^i
                                    \gamma^\mu
                                       Q_R^i
                           ] Z_\mu ,
\eea
where $Q_{L(R)}^i$ show left(right)-handed quark doublet and
$i$ shows the generation number. For ordinary quarks,
 the coefficients of couplings
$L_i$ and $R_i$ are
\bea
L_U = L_{ui} = 1 - \frac{4}{3}\sin^2\theta_w, \\
R_U = R_{ui} =   - \frac{4}{3}\sin^2\theta_w,
\eea
for up sector and
\bea
L_D = L_{di} = -1 + \frac{2}{3}\sin^2\theta_w, \\
R_D = R_{di} =      \frac{2}{3}\sin^2\theta_w,
\eea
for down sector, where $\theta_w$ is Weinberg angle.
For the vector-like quarks,
 the couplings for right-handed
fermions are the same as the left-handed ones, that is,
\bea
R^\prime_U = L_U, \\
R^\prime_D = L_D.
\eea

For the case that the fourth generation is a
vector-like doublet fermions, the neutral current are
explicitly given by the following equation.
\bea
       - \frac{g}{2 cos\theta_w}
                   &[& \pmatrix{
                 \bar{u}^1 & \bar{u}^2 & \bar{u}^3 & \bar{u}^4
                                  }_L
                     \pmatrix{
                              L_U & \  & \  & \  \cr
                              \  & L_U & \  & \  \cr
                              \  & \  & L_U & \  \cr
                              \  & \  & \  & L_U \cr
                               }
                                    \gamma^\mu
                     \pmatrix{
                              u^1 \cr u^2 \cr u^3 \cr u^4
                              }_L  \nn \\
               &+& \pmatrix{
                 \bar{u}^1 & \bar{u}^2 & \bar{u}^3 & \bar{u}^4
                                  }_R
                     \pmatrix{
                              R_U & \  & \  & \  \cr
                              \  & R_U & \  & \  \cr
                              \  & \  & R_U & \  \cr
                              \  & \  & \  & R^\prime_U \cr
                               }
                                    \gamma^\mu
                     \pmatrix{
                              u^1 \cr u^2 \cr u^3 \cr u^4
                              }_R \nn \\
               &+&  \pmatrix{
                 \bar{d}^1 & \bar{d}^2 & \bar{d}^3 & \bar{d}^4
                                  }_L
                     \pmatrix{
                              L_D & \  & \  & \  \cr
                              \  & L_D & \  & \  \cr
                              \  & \  & L_D & \  \cr
                              \  & \  & \  & L_D \cr
                               }
                                    \gamma^\mu
                     \pmatrix{
                              d^1 \cr d^2 \cr d^3 \cr d^4
                              }_L \nn \\
              &+& \pmatrix{
                 \bar{d}^1 & \bar{d}^2 & \bar{d}^3 & \bar{d}^4
                                  }_R
                     \pmatrix{
                              R_D & \  & \  & \  \cr
                              \  & R_D & \  & \  \cr
                              \  & \  & R_D & \  \cr
                              \  & \  & \  & R^\prime_D \cr
                               }
                                    \gamma^\mu
                     \pmatrix{
                              d^1 \cr d^2 \cr d^3 \cr d^4
                              }_R
                           ].
\eea
Note that only the right-handed coupling for the forth
generation of quark  differs from
those of the other three generations.
With paying attention to the difference  of the right-handed
couplings, we transform the bases from  weak eigenstate to mass
eigenstate by the unitary transformation.
\bea
       - \frac{g}{2 cos\theta_w}
                   &[& \pmatrix{
                 \bar{u} & \bar{c} & \bar{t} & \bar{u^\prime}
                                  }_L
                               L_U
                                    \gamma^\mu
                     \pmatrix{
                              u \cr c \cr t \cr u^\prime
                              }_L  \nn \\
               &+& \pmatrix{
                 \bar{u} & \bar{c} & \bar{t} & \bar{u^\prime}
                                  }_R
                   \{ R_U +
                        V_{LU}^\dagger \pmatrix{
                              0 & \  & \  & \  \cr
                              \  & 0 & \  & \  \cr
                              \  & \  & 0 & \  \cr
                              \  & \  & \  & R^\prime_U - R_U \cr
                               } V_{LU} \}
                                    \gamma^\mu
                     \pmatrix{
                              u \cr c \cr t \cr u^\prime
                              }_R \nn \\
               &+&  \pmatrix{
                 \bar{d} & \bar{s} & \bar{b} & \bar{d^\prime}
                                  }_L
                              R_U
                                    \gamma^\mu
                     \pmatrix{
                              d \cr s \cr b \cr d^\prime
                              }_L  \\
              &+& \pmatrix{
                 \bar{d} & \bar{s} & \bar{b} & \bar{d^\prime}
                                  }_R
                     \{ R_D +
                     V_{RD}^\dagger \pmatrix{
                              0 & \  & \  & \  \cr
                              \  & 0 & \  & \  \cr
                              \  & \  & 0 & \  \cr
                              \  & \  & \  & R^\prime_D - R_D \cr
                               } V_{RD} \}
                                    \gamma^\mu
                     \pmatrix{
                              d \cr s \cr b \cr d^\prime
                              }_R
                           ].  \nn
\eea
In the mass eigenstates,
the left-handed current dose not change but the right-handed current
 changes as following;
\bea
g_R^U = - \frac{g}{2c_w}\{ R_U
              + \pmatrix{
                  \mid V_{14}\mid^2 & V_{14}^\ast V_{42}
                  & V_{14}^\ast V_{43} & V_{14}^\ast V_{44}
                  \cr
                  V_{24}^\ast V_{41} & \mid V_{24}\mid^2
                  & V_{24}^\ast V_{43} & V_{24}^\ast V_{44}
                  \cr
                  V_{34}^\ast V_{41} & V_{34}^\ast V_{42}
                  & \mid V_{34}\mid^2 & V_{34}^\ast V_{44}
                  \cr
                   V_{44}^\ast V_{41} & V_{44}^\ast V_{42}
                  & V_{44}^\ast V_{43} & \mid V_{44}\mid^2
                         }_{RU} \}, \\
g_R^D = - \frac{g}{2c_w}\{ R_D
              - \pmatrix{
                  \mid V_{14}\mid^2 & V_{14}^\ast V_{42}
                  & V_{14}^\ast V_{43} & V_{14}^\ast V_{44}
                  \cr
                  V_{24}^\ast V_{41} & \mid V_{24}\mid^2
                  & V_{24}^\ast V_{43} & V_{24}^\ast V_{44}
                  \cr
                  V_{34}^\ast V_{41} & V_{34}^\ast V_{42}
                  & \mid V_{34}\mid^2 & V_{34}^\ast V_{44}
                  \cr
                   V_{44}^\ast V_{41} & V_{44}^\ast V_{42}
                  & V_{44}^\ast V_{43} & \mid V_{44}\mid^2
                         }_{RD} \}.
\eea
As shown in the above equations, there are FCNC among four
generations in general.
Because  FCNC among light quarks  like $Z \to ds$ are
strongly constrained by the experiments,
we simply suppose that there are no couplings which produce FCNC among
the light quarks in this model. The model which  corresponds to the
simple assumption will be discussed in section 2.
Before going into the  model that the quark of the forth
generation is vector-like doublet,
 let us briefly study the models
that the quark of the third generation is a vector-like
doublet \cite{ple} and show why those models are not
consistent with the experiments.
In the case that the third generation is a vector like
doublet and there is no more generations,
 the right-handed coupling of $Z$ boson is
\bea
g_R^{(U,D)} = - \frac{g}{2c_w}\{ R_{(U,D)}
              \pm \pmatrix{
                  \mid V_{13}\mid^2 & V_{13}^\ast V_{32}
                  & V_{13}^\ast V_{33}
                  \cr
                  V_{23}^\ast V_{31} & \mid V_{23}\mid^2
                  & V_{23}^\ast V_{33}
                  \cr
                  V_{33}^\ast V_{31} & V_{33}^\ast V_{32}
                  & \mid V_{33}\mid^2
                         }_{R(U,D)} \}.
\eea
{}From the unitarity of matrix $V_R$, we obtain the
following constraint for the mixing of right-handed quarks.
\bea
\mid V_{13}\mid^2 + \mid V_{23}\mid^2 +\mid V_{33}\mid^2 = 1.
\label{uni3}
\eea
In order to avoid the FCNC among light quarks,
the values of $V_{13}$ and $V_{23}$ must be very
small. Then, by using the unitarity condition
eq.(\ref{uni3}), $\mid V_{33}\mid \sim 1$.
This means that the  the right handed coupling of b quark
must be almost the same as that of  the left-handed.
However,in that case, the constraint from $A_{FB}$ can not
be satisfied.
 If the right handed coupling of b quark
is the same as that of  the left-handed one, there is no neutral axial
vector current in $Z bb$ vertex and  the forward-backward
asymmetry, $A_{FB}^b$, must disappear.

Even if we add more vector-like doublet quarks, the
situation is not changed. To show this, we consider
 the case that the third generation and the forth
generation are vector-like
doublets.
 The right-handed coupling of $Z$ boson to down
type quarks is given by,
(Here we assume that there is no FCNC among the
light quarks.)
\bea
g_R^{(D)} = - \frac{g}{2c_w}\{ R_{(D)}
              -  \pmatrix{
                  0 & 0
                  & 0 & 0
                  \cr
                  0 & 0
                  & 0 & 0
                  \cr
                  0 & 0
                  & \mid V_{33}\mid^2 + \mid V_{34}\mid^2
                  & V_{33}^\ast V_{34} + V_{34}^\ast V_{44}
                  \cr
                   0 & 0
                  & V_{34}^\ast V_{33} + V_{44}^\ast V_{43}
                  & \mid V_{34}\mid^2 + \mid V_{44}\mid^2
                         }_{R(D)} \},
\eea
where we put $V_{13} \sim V_{23} \sim V_{14} \sim V_{24}
\sim 0$.
{}From the unitarity of the matrix $V_R$,
\bea
\mid V_{33}\mid^2 +\mid V_{34}\mid^2 = 1.
\label{uni4}
\eea
Again eq.(\ref{uni4}) show that $Zbb$ coupling must be
vector type
and it contradicts
with  the ${A^b}_{FB}$ because of the
same reason as the first  case.

Hence, we consider the model in which only the forth generation
is a vector-like doublet quarks.

\section{A Simple Model}
There are mixings among the heavy quarks but are not
among the light quarks to avoid the problems of FCNC from $Z$
couplings.  The neutral couplings of the ordinary quarks are
\bea
g_L^u = g_L^c &=& L_U,  \\
g_L^d = g_L^s = g_L^b &=& L_D,  \\
g_R^u &=& R_U,  \\
g_R^c &=& R_U + {V_{RU}^{42}}^2,  \\
g_R^d = g_R^s &=& R_D, \\
g_R^b &=& R_D - {V_{RD}^{43}}^2,
\eea
where $V_{RU}^{42}$ and $ V_{RD}^{43}$ are the component of unitary
matrix $V_R$. The difference of the sign of the $V^2$
between $g_R^c$ and $g_R^b$ come from one of the isospin.
The experiment favor positive
contribution for $R_b$ and negative contribution for$R_c$.
Then the partial width $R_b$ and $R_c$ is
\bea
R_b^{tree} &\sim& \frac{{g_L^b}^2 + {g_R^b}^2}
             {{g_L^u}^2 + {g_L^c}^2 + {g_L^d}^2 + {g_L^s}^2 +
                  {g_L^b}^2 + {g_R^u}^2 + {g_R^c}^2 +
                  {g_R^d}^2 + {g_R^s}^2 + {g_R^b}^2 } \nn \\
           &\sim& \frac{ L_D^2 + (R_D - {V_{RD}^{43}}^2)^2}
             { 2 L_U^2 + 3 L_D^2 + R_U^2 + 2 R_D^2 +
               (R_U + {V_{RU}^{42}}^2)^2 + (R_D - {V_{RD}^{43}}^2)^2},
\eea
\bea
R_c^{tree} &\sim& \frac{{g_L^c}^2 + {g_R^c}^2}
             {{g_L^u}^2 + {g_L^c}^2 + {g_L^d}^2 + {g_L^s}^2 +
                  {g_L^b}^2 + {g_R^u}^2 + {g_R^c}^2 +
                  {g_R^d}^2 + {g_R^s}^2 + {g_R^b}^2 } \nn \\
           &\sim& \frac{ L_U^2 + (R_U + {V_{RU}^{42}}^2)^2}
             { 2 L_U^2 + 3 L_D^2 + R_U^2 + 2 R_D^2 +
               (R_U + {V_{RU}^{42}}^2)^2 + (R_D - {V_{RD}^{43}}^2)^2}.
\eea
Because of the minus sign of ${V_{RD}^{43}}^2$ in the
$g_R^b$, unless the values is larger than $2 \times R_D$,
the contribution from $V^2$ reduce the $R_b$.
If ${V_{RD}^{43}}^2$ is larger than $2 \times R_D$ and
$R_U + {V_{RU}^{42}}^2$ become smaller, $R_b$ may be
enhanced and $R_c$ may become smaller.
While, we must also examine the allowed region of these parameter
$V_{RU}$ and $V_{RD}$ for forward-backward asymmetry
$A_{FB}^c$ and $A_{FB}^b$. The values of the experiment are
$A_{FB}^{0,b} = 0.0997\pm0.0031$ and
$A_{FB}^{0,c} = 0.0729\pm0.0058$\cite{hagi}.
The prediction of SM are $A_{FB}^{bSM} = 0.1039$ and
$A_{FB}^{cSM} = 0.744$ with $m_t = 175GeV$ and Higgs mass
$m_H = 1TeV$. In Fig.1 and Fig.2, we plot the
behavior of $\delta A_{FB}^b$ and $\delta A_{FB}^c$ which
are shifting the SM result due to $V_{RD}$ and
$V_{RU}$.
The formalism of the forward-backward asymmetry is
\bea
A_{FB}^f = \frac{\sigma_F - \sigma_B}{\sigma_F + \sigma_B}
\eea
where $\sigma_F$ and $\sigma_B$ are the cross sections for
the outgoing fermion, f, to go forwards or backwards
relative to the direction of the incoming electron. When the
center of mass, $\sqrt{s}$, is about the mass $M_Z$,
the forward-backward asymmetry, $A_{FB}$ is shown by the
following,
\bea
A_{FB}^f = \frac{3}{4} A_e A_f,
\eea
where,
\bea
A_f = \frac{2 (g_L + g_R)(g_L - g_R)}{(g_L + g_R)^2 + (g_L - g_R)^2}.
\eea
It is interesting that the value of forward-backward
asymmetry do not change by replacing the $g_R$ with $- g_R$.
$R_b$ and $R_c$ have also the same feature.
Hence, we expect that $R_b$ and $A_{FB}^b$ with
${V_{RD}^{43}}^2  = 2 \times R_D$ is same the values with
${V_{RD}^{43}}^2  =  0$.
The range of shifting of the right-handed Z couplings must satisfy
the constraint from the forward-backward asymmetry.
The contribution from the shifting of the neutral couplings
to the forward-backward asymmetry is
\bea
\delta A_{FB}^b = - \frac{3}{4} A_e ( 1 + A_b )
               \frac{4 R_D ({V_{RD}^{43}}^2 - 2 R_D)}
                      {(L_D + R_D)^2 + (L_D - R_D)^2},\\
\delta A_{FB}^c = - \frac{3}{4} A_e ( 1 + A_c )
               \frac{4 R_U {V_{RU}^{42}}^2}
                      {(L_U + R_U)^2 + (L_U - R_U)^2}.
\eea
Here, we consider the contribution to $A_{FB}^b$ is the
value for the shift of ${V_{RD}^{43}}^2$ from $2 \times R_D$.
In Fig.1, the behavior of the $\delta A_{FB}^b$ is shown.
{}From the experimental data, the allowed region of $\delta
A_{FB}^b$ must be from - 0.0011 to - 0.0073. Then, the value
of ${V_{RD}^{43}}^2$ must be smaller than about 0.38. While, in Fig.2,
the behavior of $\delta A_{FB}^c$ is shown.
The allowed region of $\delta
A_{FB}^c$ must be 0.0043 $\sim$ - 0.0073. Then, the upper
bound for the value
of ${V_{RU}^{24}}^2$ is about 0.033. For the region
satisfying the constraint from forward-backward asymmetry,
we plot $R_b$ and $R_c$ in Fig.3 and Fig.4 in the following
case of ${V_{RU}^{24}}^2$. (1) ${V_{RU}^{24}}^2 = 0.01$, (2)
0.02, (3)0.03. Here, we used the prediction of SM with $m_t = 175GeV$
and add the contribution from shifting the right-hand
coupling to its values. There is a
region which is satisfying both conditions for $R_b$ and $A_{FB}^b$
but not for $R_c$, $A_{FB}^b$ and $A_{FB}^c$. With
${V_{RD}^{34}}^2 = 0.36$, the prediction of $R_b$ reach to
the center values of experiment. The figures
show that
The effect of vector-like doublet as the forth generation
may explain the difference of $R_b$ between the experiment
and SM
but can not explain the difference of $R_c$.

On the other hand the charged current and the KM matrix are
\bea
{\cal L}_R^{cc} = - \frac{g}{\sqrt{2}}
                   \pmatrix{
                 \bar{u} & \bar{c} & \bar{t} & \bar{u^\prime}
                                  }_R
                     V_{RU}^\dagger \pmatrix{
                              0 & \  & \  & \  \cr
                              \  & 0 & \  & \  \cr
                              \  & \  & 0 & \  \cr
                              \  & \  & \  & 1 \cr
                               } V_{RD}
                                    \gamma^\mu
                     \pmatrix{
                              d \cr s \cr b \cr d^\prime
                              }_R W_\mu^\dagger.
\eea
For example, the KM matrix becomes
\bea
V_{KM}^{R} = \pmatrix{
              0 & 0 & 0 & 0 \cr
              0 & 0 &
 {V_{RU}^{24}}^\ast V_{RD}^{43} & {V_{RU}^{24}}^\ast V_{RD}^{44}
                                         \cr
              0 & 0 &
 {V_{RU}^{34}}^\ast V_{RD}^{43} & {V_{RU}^{34}}^\ast V_{RD}^{44}
                                         \cr
              0 & 0 &
 {V_{RU}^{44}}^\ast V_{RD}^{43} &  {V_{RU}^{44}}^\ast V_{RD}^{44}
                         }.
\eea
Note that this matrix is not unitary. In this paper, we did
not discussed the effect from the right hand KM matrix. By
upper discussion, we may be able to examine the value of
component of the matrix. We must consider the contribution from the process
like $b \to s \gamma$.

\section{Flavor Mixing}
We must examine whether such large mixing between b quark
and the fourth down type quark is a realistic model.
We consider the simple vector-like doublet
model which are assumed to mix only with the
third generation\cite{han,lav}. Without loss of generality, the quark mass
terms are
\bea
   \pmatrix{
           \bar{u}^3 & \bar{u}^4
                                  }_L
           \pmatrix{
           m_u & 0 \cr
           J e^{ix} & M_V
                    }
                     \pmatrix{
                              u^3 \cr u^4
                              }_R
+    \pmatrix{
           \bar{d}^3 & \bar{d}^4
                                  }_L
           \pmatrix{
           m_d & 0 \cr
           K & M_V
                    }
                     \pmatrix{
                              d^3 \cr d^4
                              }_R  + h.c.,
\eea
where $M_V$ is a mass of vector-like quarks, $J, K$ show
mixing mass. By using the unitary transformation the mass
matrix is diagonalized as following,
\bea
\pmatrix{
         c_{LU} & - s_{LU} e^{-ix} \cr
         s_{LU}e^{ix} & c_{LU}
         }
         \pmatrix{
            m_u & 0 \cr
               J  e^{ix} & M_V
                    }
                    \pmatrix{
                         c_{RU} & - s_{RU}e^{-ix} \cr
                         s_{RU}e^{ix} & c_{RU}
                            } =
         \pmatrix{
                 m_t & 0 \cr
                     0 & M_u
                    } \\[5mm]
\pmatrix{
         c_{LD} & - s_{LD} \cr
         s_{LD} & c_{LD}
         }
         \pmatrix{
            m_d & 0 \cr
                K & M_V
                    }
                    \pmatrix{
                         c_{RD} & - s_{RD} \cr
                         s_{RD} & c_{RD}
                            } =
         \pmatrix{
                 m_b & 0 \cr
                     0 & M_d
                    },
\eea
where $c_{XX} = cos\theta_{XX}$ and $s_{XX} =
sin\theta_{XX}$ and the $\theta_{XX}$ show the mixing angle.
The masses of eigen states are
\bea
m_t^2 &=& \frac{1}{2}[ m_u^2 + M_V^2 + J^2
          - \sqrt{(m_u^2 + M_V^2 + J^2)^2
                   - 4 m_u^2 M_V^2} ],\\
M_u^2 &=& \frac{1}{2}[ m_u^2 + M_V^2 + J^2
          + \sqrt{(m_u^2 + M_V^2 + J^2)^2
                   - 4 m_u^2 M_V^2} ],\\
m_b^2 &=& \frac{1}{2}[ m_d^2 + M_V^2 + K^2
          - \sqrt{(m_d^2 + M_V^2 + K^2)^2
                   - 4 m_d^2 M_V^2}], \\
M_d^2 &=& \frac{1}{2}[ m_d^2 + M_V^2 + K^2
          + \sqrt{(m_d^2 + M_V^2 + K^2)^2
                   - 4 m_d^2 M_V^2}].
\eea
The mixing just corresponding to $V_{34}^2$ are
\bea
s_{RU}^2 &=& \frac{(m_u^2 - M_V^2 + J^2)
                 (m_u^2 + J^2 - m_t^2) + 2 J^2 M_V^2}
           {(m_u^2 - M_V^2 + J^2)^2
             - 4 J^2 M_V^2},\\
s_{RU} &=& \frac{M_V}{m_t} s_{LU},
\eea
\bea
s_{RD}^2 &=& \frac{(m_d^2 - M_V^2 + K^2)
                 (m_d^2 + K^2 - m_b^2) + 2 K^2 M_V^2}
           {(m_d^2 - M_V^2 + J^2)^2
             - 4 K^2 M_V^2,}\\
s_{RD} &=& \frac{M_V}{m_b} s_{LD}.
\label{sl}
\eea
In Fig. 5, we plot $s_{RD}^2$ as a function of $K$ in
the several case of $M_V$, 1TeV, and 2TeV and 3TeV. Here, we
put $m_b = 5GeV$. The
Figure shows that there is a region to produce
large mixing.  $V_{34}^2 \sim s_{RD}^2 \sim 0.35$.
It is interesting that even if the mixing for right-hand quark is
very large, the left-hand one is kept very small because
of eq.(\ref{sl}). Hence, the CKM matrix for left-handed
quarks will not make a large different from SM. Since we
assume a very large
vector-like quark mass, the dependence to Oblique correction\cite{pes}
will be small by means of decoupling\cite{lav}.

\section{Conclusion}
We discussed the effect of vector-like doublet quarks to Z
partial decay width. Under consideration of some constraint
from FCNC for light quarks and forward-backward asymmetry,
only model of vector-like quarks as the forth generation is allowed.
We find that this model may explain the difference of $R_b$
between the experiment and the prediction of SM but not do
for $R_c$. We will have to consider the other
contribution to $R_c$.

In this work, we could not do sufficiently discussion for
flavor physics. If we apply this model, the contribution to
$b\rightarrow s \gamma$ must be calculated.

\bigskip
\bigskip

{\noindent {\Large{\bf Acknowledgement}}} \\
I would like to thank Dr.T.Morozumi and L.T.Handoko for
helpful discussions.

\newpage
\begin{center}
{\Large{\bf Figure Captions}}
\end{center}
\begin{itemize}
\item {\bf Figure 1} Plotting the $\delta A_{FB}^b$ as a function of
${V_{R34}^D}^2$.

\item {\bf Figure 2} Plotting the $\delta A_{FB}^c$ as a function of
${V_{R24}^U}^2$.

\item {\bf Figure 3} $R_b$ as a function of ${V_{R34}^D}^2$
for following values for ${V_{R24}^U}^2$. (1) 0.01 with a
thinline, (2) 0.02 with a thickline, (3) 0.03 with s dashed thinline.

\item {\bf Figure 4} $R_c$ as a function of ${V_{R34}^D}^2$
for following values for ${V_{R24}^U}^2$. (1) 0.01 with a
thinline, (2) 0.02 with a thickline, (3) 0.03 with s dashed thinline.

\item {\bf Figure 5}$s_{RD}^2$ as a function of mass mixing, $K$, for
following mass $M_V$. (a) 1 TeV with a dashed thinline, (b)
2 TeV with a dashed thickline and (c) 3 TeV with a thickline.
\end{itemize}
\newpage

\begin{figure}[htb]
\setlength{\unitlength}{0.240900pt}
\begin{picture}(1500,900)(0,0)
\tenrm
\thicklines \path(220,113)(240,113)
\thicklines \path(1436,113)(1416,113)
\put(198,113){\makebox(0,0)[r]{-0.008}}
\thicklines \path(220,198)(240,198)
\thicklines \path(1436,198)(1416,198)
\thicklines \path(220,283)(240,283)
\thicklines \path(1436,283)(1416,283)
\put(198,283){\makebox(0,0)[r]{-0.006}}
\thicklines \path(220,368)(240,368)
\thicklines \path(1436,368)(1416,368)
\thicklines \path(220,453)(240,453)
\thicklines \path(1436,453)(1416,453)
\put(198,453){\makebox(0,0)[r]{-0.004}}
\thicklines \path(220,537)(240,537)
\thicklines \path(1436,537)(1416,537)
\thicklines \path(220,622)(240,622)
\thicklines \path(1436,622)(1416,622)
\put(198,622){\makebox(0,0)[r]{-0.002}}
\thicklines \path(220,707)(240,707)
\thicklines \path(1436,707)(1416,707)
\thicklines \path(220,792)(240,792)
\thicklines \path(1436,792)(1416,792)
\put(198,792){\makebox(0,0)[r]{0}}
\thicklines \path(220,877)(240,877)
\thicklines \path(1436,877)(1416,877)
\thicklines \path(220,113)(220,133)
\thicklines \path(220,877)(220,857)
\put(220,68){\makebox(0,0){0.3}}
\thicklines \path(382,113)(382,133)
\thicklines \path(382,877)(382,857)
\put(382,68){\makebox(0,0){0.32}}
\thicklines \path(544,113)(544,133)
\thicklines \path(544,877)(544,857)
\put(544,68){\makebox(0,0){0.34}}
\thicklines \path(706,113)(706,133)
\thicklines \path(706,877)(706,857)
\put(706,68){\makebox(0,0){0.36}}
\thicklines \path(869,113)(869,133)
\thicklines \path(869,877)(869,857)
\put(869,68){\makebox(0,0){0.38}}
\thicklines \path(1031,113)(1031,133)
\thicklines \path(1031,877)(1031,857)
\put(1031,68){\makebox(0,0){0.4}}
\thicklines \path(1193,113)(1193,133)
\thicklines \path(1193,877)(1193,857)
\put(1193,68){\makebox(0,0){0.42}}
\thicklines \path(1355,113)(1355,133)
\thicklines \path(1355,877)(1355,857)
\put(1355,68){\makebox(0,0){0.44}}
\thicklines \path(220,113)(1436,113)(1436,877)(220,877)(220,113)
\put(45,505){\makebox(0,0)[l]{\shortstack{$ \delta A_{FB}^b$}}}
\put(828,23){\makebox(0,0){$ {V_{R34}^D}^2 $}}
\thinlines \path(220,824)(220,824)(271,794)(321,764)
(372,735)(423,705)(473,676)(524,646)(575,616)(625,587)
(676,557)(727,528)(777,498)(828,468)(879,439)(929,409)
(980,379)(1031,350)(1081,320)(1132,291)(1183,261)(1233,231)
(1284,202)(1335,172)(1385,143)(1436,113)
\end{picture}

\caption{~}
\end{figure}

\begin{figure}[htb]
\setlength{\unitlength}{0.240900pt}
\begin{picture}(1500,900)(0,0)
\tenrm
\thicklines \path(220,113)(240,113)
\thicklines \path(1436,113)(1416,113)
\put(198,113){\makebox(0,0)[r]{0}}
\thicklines \path(220,222)(240,222)
\thicklines \path(1436,222)(1416,222)
\put(198,222){\makebox(0,0)[r]{0.001}}
\thicklines \path(220,331)(240,331)
\thicklines \path(1436,331)(1416,331)
\put(198,331){\makebox(0,0)[r]{0.002}}
\thicklines \path(220,440)(240,440)
\thicklines \path(1436,440)(1416,440)
\put(198,440){\makebox(0,0)[r]{0.003}}
\thicklines \path(220,550)(240,550)
\thicklines \path(1436,550)(1416,550)
\put(198,550){\makebox(0,0)[r]{0.004}}
\thicklines \path(220,659)(240,659)
\thicklines \path(1436,659)(1416,659)
\put(198,659){\makebox(0,0)[r]{0.005}}
\thicklines \path(220,768)(240,768)
\thicklines \path(1436,768)(1416,768)
\put(198,768){\makebox(0,0)[r]{0.006}}
\thicklines \path(220,877)(240,877)
\thicklines \path(1436,877)(1416,877)
\put(198,877){\makebox(0,0)[r]{0.007}}
\thicklines \path(220,113)(220,133)
\thicklines \path(220,877)(220,857)
\put(220,68){\makebox(0,0){0}}
\thicklines \path(342,113)(342,133)
\thicklines \path(342,877)(342,857)
\put(342,68){\makebox(0,0){0.005}}
\thicklines \path(463,113)(463,133)
\thicklines \path(463,877)(463,857)
\put(463,68){\makebox(0,0){0.01}}
\thicklines \path(585,113)(585,133)
\thicklines \path(585,877)(585,857)
\put(585,68){\makebox(0,0){0.015}}
\thicklines \path(706,113)(706,133)
\thicklines \path(706,877)(706,857)
\put(706,68){\makebox(0,0){0.02}}
\thicklines \path(828,113)(828,133)
\thicklines \path(828,877)(828,857)
\put(828,68){\makebox(0,0){0.025}}
\thicklines \path(950,113)(950,133)
\thicklines \path(950,877)(950,857)
\put(950,68){\makebox(0,0){0.03}}
\thicklines \path(1071,113)(1071,133)
\thicklines \path(1071,877)(1071,857)
\put(1071,68){\makebox(0,0){0.035}}
\thicklines \path(1193,113)(1193,133)
\thicklines \path(1193,877)(1193,857)
\put(1193,68){\makebox(0,0){0.04}}
\thicklines \path(1314,113)(1314,133)
\thicklines \path(1314,877)(1314,857)
\put(1314,68){\makebox(0,0){0.045}}
\thicklines \path(1436,113)(1436,133)
\thicklines \path(1436,877)(1436,857)
\put(1436,68){\makebox(0,0){0.05}}
\thicklines \path(220,113)(1436,113)(1436,877)(220,877)(220,113)
\put(45,495){\makebox(0,0)[l]{\shortstack{$ \delta A_{FB}^c$}}}
\put(828,23){\makebox(0,0){$ {V_{R24}^U}^2$}}
\thinlines \path(220,113)(220,113)(271,143)(321,173)
(372,204)(423,234)(473,264)(524,294)(575,324)(625,355)
(676,385)(727,415)(777,445)(828,475)(879,505)(929,536)
(980,566)(1031,596)(1081,626)(1132,656)(1183,687)
(1233,717)(1284,747)(1335,777)(1385,807)(1436,838)
\end{picture}

\caption{~}
\end{figure}

\begin{figure}[htb]
\setlength{\unitlength}{0.240900pt}
\begin{picture}(1500,900)(0,0)
\tenrm
\thicklines \path(220,189)(240,189)
\thicklines \path(1436,189)(1416,189)
\put(198,189){\makebox(0,0)[r]{0.215}}
\thicklines \path(220,380)(240,380)
\thicklines \path(1436,380)(1416,380)
\put(198,380){\makebox(0,0)[r]{0.22}}
\thicklines \path(220,571)(240,571)
\thicklines \path(1436,571)(1416,571)
\put(198,571){\makebox(0,0)[r]{0.225}}
\thicklines \path(220,762)(240,762)
\thicklines \path(1436,762)(1416,762)
\put(198,762){\makebox(0,0)[r]{0.23}}
\thicklines \path(220,113)(220,133)
\thicklines \path(220,877)(220,857)
\put(220,68){\makebox(0,0){0.25}}
\thicklines \path(524,113)(524,133)
\thicklines \path(524,877)(524,857)
\put(524,68){\makebox(0,0){0.3}}
\thicklines \path(828,113)(828,133)
\thicklines \path(828,877)(828,857)
\put(828,68){\makebox(0,0){0.35}}
\thicklines \path(1132,113)(1132,133)
\thicklines \path(1132,877)(1132,857)
\put(1132,68){\makebox(0,0){0.4}}
\thicklines \path(1436,113)(1436,133)
\thicklines \path(1436,877)(1436,857)
\put(1436,68){\makebox(0,0){0.45}}
\thicklines \path(220,113)(1436,113)(1436,877)(220,877)(220,113)
\put(45,495){\makebox(0,0)[l]{\shortstack{$ R_b $}}}
\put(828,23){\makebox(0,0){$ {V_{R34}^D}^2$}}
\thinlines \path(295,113)(321,122)(372,139)(423,158)
(473,178)(524,199)(575,222)(625,246)(676,271)(727,297)
(777,324)(828,353)(879,382)(929,413)(980,445)(1031,478)
(1081,513)(1132,548)(1183,585)(1233,622)(1284,661)
(1335,701)(1385,742)(1436,784)
\Thicklines \path(249,113)(271,119)(321,136)(372,153)
(423,172)(473,192)(524,214)(575,236)(625,260)(676,285)
(727,311)(777,339)(828,367)(879,397)(929,428)(980,460)
(1031,493)(1081,527)(1132,563)(1183,599)(1233,637)(1284,676)
(1335,716)(1385,757)(1436,799)
\thinlines \dashline[-10]{25}(220,118)(220,118)(271,133)
(321,150)(372,167)(423,186)(473,206)(524,228)(575,250)
(625,274)(676,299)(727,325)(777,353)(828,381)(879,411)
(929,442)(980,474)(1031,507)(1081,542)(1132,577)(1183,614)
(1233,652)(1284,690)(1335,730)(1385,772)(1436,814)
\end{picture}

\caption{~}
\end{figure}

\begin{figure}[htb]
\setlength{\unitlength}{0.240900pt}
\begin{picture}(1500,900)(0,0)
\tenrm
\thicklines \path(220,113)(240,113)
\thicklines \path(1436,113)(1416,113)
\put(198,113){\makebox(0,0)[r]{0.16}}
\thicklines \path(220,189)(240,189)
\thicklines \path(1436,189)(1416,189)
\put(198,189){\makebox(0,0)[r]{0.162}}
\thicklines \path(220,266)(240,266)
\thicklines \path(1436,266)(1416,266)
\put(198,266){\makebox(0,0)[r]{0.164}}
\thicklines \path(220,342)(240,342)
\thicklines \path(1436,342)(1416,342)
\put(198,342){\makebox(0,0)[r]{0.166}}
\thicklines \path(220,419)(240,419)
\thicklines \path(1436,419)(1416,419)
\put(198,419){\makebox(0,0)[r]{0.168}}
\thicklines \path(220,495)(240,495)
\thicklines \path(1436,495)(1416,495)
\put(198,495){\makebox(0,0)[r]{0.17}}
\thicklines \path(220,571)(240,571)
\thicklines \path(1436,571)(1416,571)
\put(198,571){\makebox(0,0)[r]{0.172}}
\thicklines \path(220,648)(240,648)
\thicklines \path(1436,648)(1416,648)
\put(198,648){\makebox(0,0)[r]{0.174}}
\thicklines \path(220,724)(240,724)
\thicklines \path(1436,724)(1416,724)
\put(198,724){\makebox(0,0)[r]{0.176}}
\thicklines \path(220,801)(240,801)
\thicklines \path(1436,801)(1416,801)
\put(198,801){\makebox(0,0)[r]{0.178}}
\thicklines \path(220,877)(240,877)
\thicklines \path(1436,877)(1416,877)
\put(198,877){\makebox(0,0)[r]{0.18}}
\thicklines \path(220,113)(220,133)
\thicklines \path(220,877)(220,857)
\put(220,68){\makebox(0,0){0.25}}
\thicklines \path(524,113)(524,133)
\thicklines \path(524,877)(524,857)
\put(524,68){\makebox(0,0){0.3}}
\thicklines \path(828,113)(828,133)
\thicklines \path(828,877)(828,857)
\put(828,68){\makebox(0,0){0.35}}
\thicklines \path(1132,113)(1132,133)
\thicklines \path(1132,877)(1132,857)
\put(1132,68){\makebox(0,0){0.4}}
\thicklines \path(1436,113)(1436,133)
\thicklines \path(1436,877)(1436,857)
\put(1436,68){\makebox(0,0){0.45}}
\thicklines \path(220,113)(1436,113)(1436,877)(220,877)(220,113)
\put(45,495){\makebox(0,0)[l]{\shortstack{$ R_c $}}}
\put(828,23){\makebox(0,0){$ {V_{R34}^D}^2$}}
\thinlines \path(220,521)(220,521)(271,518)(321,515)(372,511)
(423,507)(473,502)(524,498)(575,493)(625,488)(676,482)(727,477)
(777,471)(828,465)(879,458)(929,451)(980,444)(1031,437)(1081,430)
(1132,422)(1183,414)(1233,406)(1284,398)(1335,389)(1385,380)(1436,371)
\Thicklines \path(220,466)(220,466)(271,463)(321,459)(372,455)
(423,451)(473,447)(524,442)(575,437)(625,432)(676,427)(727,421)
(777,415)(828,409)(879,403)(929,396)(980,389)(1031,382)
(1081,375)(1132,367)(1183,359)(1233,351)(1284,343)(1335,334)
(1385,326)(1436,316)
\thinlines \dashline[-10]{25}(220,412)(220,412)(271,409)
(321,405)(372,401)(423,397)(473,393)(524,389)(575,384)
(625,379)(676,373)(727,368)(777,362)(828,356)(879,350)
(929,343)(980,336)(1031,329)(1081,322)(1132,314)(1183,306)
(1233,298)(1284,290)(1335,281)(1385,273)(1436,264)
\end{picture}

\caption{~}
\end{figure}

\begin{figure}[htb]
\setlength{\unitlength}{0.240900pt}
\begin{picture}(1500,900)(0,0)
\tenrm
\thicklines \path(220,113)(240,113)
\thicklines \path(1436,113)(1416,113)
\put(198,113){\makebox(0,0)[r]{0}}
\thicklines \path(220,189)(240,189)
\thicklines \path(1436,189)(1416,189)
\thicklines \path(220,266)(240,266)
\thicklines \path(1436,266)(1416,266)
\put(198,266){\makebox(0,0)[r]{0.1}}
\thicklines \path(220,342)(240,342)
\thicklines \path(1436,342)(1416,342)
\thicklines \path(220,419)(240,419)
\thicklines \path(1436,419)(1416,419)
\put(198,419){\makebox(0,0)[r]{0.2}}
\thicklines \path(220,495)(240,495)
\thicklines \path(1436,495)(1416,495)
\thicklines \path(220,571)(240,571)
\thicklines \path(1436,571)(1416,571)
\put(198,571){\makebox(0,0)[r]{0.3}}
\thicklines \path(220,648)(240,648)
\thicklines \path(1436,648)(1416,648)
\thicklines \path(220,724)(240,724)
\thicklines \path(1436,724)(1416,724)
\put(198,724){\makebox(0,0)[r]{0.4}}
\thicklines \path(220,801)(240,801)
\thicklines \path(1436,801)(1416,801)
\thicklines \path(220,877)(240,877)
\thicklines \path(1436,877)(1416,877)
\put(198,877){\makebox(0,0)[r]{0.5}}
\thicklines \path(220,113)(220,133)
\thicklines \path(220,877)(220,857)
\put(220,68){\makebox(0,0){500}}
\thicklines \path(524,113)(524,133)
\thicklines \path(524,877)(524,857)
\put(524,68){\makebox(0,0){1000}}
\thicklines \path(828,113)(828,133)
\thicklines \path(828,877)(828,857)
\put(828,68){\makebox(0,0){1500}}
\thicklines \path(1132,113)(1132,133)
\thicklines \path(1132,877)(1132,857)
\put(1132,68){\makebox(0,0){2000}}
\thicklines \path(1436,113)(1436,133)
\thicklines \path(1436,877)(1436,857)
\put(1436,68){\makebox(0,0){2500}}
\thicklines \path(220,113)(1436,113)(1436,877)(220,877)(220,113)
\put(45,495){\makebox(0,0)[l]{\shortstack{$ s_{RD}^2 $}}}
\put(828,23){\makebox(0,0){$ K $}}
\Thicklines \path(220,154)(220,154)(271,169)(321,185)
(372,203)(423,222)(473,243)(524,266)(575,289)(625,314)
(676,339)(727,365)(777,392)(828,419)(879,446)(929,473)
(980,501)(1031,528)(1081,556)(1132,583)(1183,610)(1233,637)
(1284,663)(1335,689)(1385,714)(1436,739)
\Thicklines  \dashline[-10]{25}(220,203)(271,233)(321,266)
(372,301)(423,339)(473,378)(524,419)(575,460)(625,501)(676,542)
(727,583)(777,624)(828,663)(879,702)(929,739)(980,776)(1031,811)
(1081,845)(1132,877)
\thinlines  \dashline[-10]{25}(220,419)(220,419)(271,501)
(321,583)(372,663)(423,739)(473,811)(524,877)
\end{picture}

\caption{~}
\end{figure}

\end{document}